\documentclass{ws-procs9x6}

\newcommand{\bk}{{\mathbf k}}
\newcommand{\bq}{{\mathbf q}}
\newcommand{\bzero}{{\mathbf 0}}

\begin{document}

\title{Mode Softening near the critical point within effective
  approaches to QCD}

\author{H.~FUJII}

\address{Institute of Physics, University of Tokyo, Komaba, Tokyo 153-8902, 
Japan\\
E-mail: hfujii@phys.c.u-tokyo.ac.jp}

\author{M.~OHTANI}

\address{Radiation Laboratory, RIKEN, Wako, Saitama 351-0198, Japan\\
E-mail: ohtani@rarfaxp.riken.jp}  

\maketitle

\abstracts{
We study the soft mode along
the critical line in the phase diagram with the tricritical point, 
using the Nambu--Jona-Lasinio model.
At the critical point with finite quark mass, the ordering density becomes 
a linear combination of the scalar, quark number and energy densities,
and their susceptibilities diverge with the same exponent.
Based on the conservation law,
it is argued that
the divergent susuceptibility of a conserved density must be
accompanied by a critically--slowing hydrodynamic mode.
The shift of the soft mode from the sigma meson to the hydrodynamic
mode occurs at the tricritical point on the critical line.
}

\section{Introduction}
The phases of QCD have been explored by investigating
the behavior of 
the quark condensate, the Polyakov loop and the color superconducting
gap as functions of the temperature ($T$)
and the quark chemical potential ($\mu$).
Among various possibilities on the phase structure,
existence of a critical point (CP)
as an endpoint of the first--order phase boundary
has been theoretically suggested and discussed 
in the literature\cite{MAS04,RS98,RS99}.

\begin{figure}[tb]
\centerline{\epsfxsize=0.5\textwidth\epsfbox{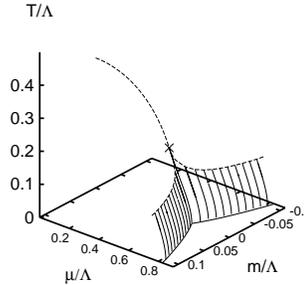}}
\caption{Phase diagram of the NJL model. The critical line
is drawn in a dashed line. The first--order phase boundary 
forms a surface shown by hatch.
The TCP is indicated by $\times$.} 
\label{fig:phase}
\end{figure}

When we extend the phase space by taking the  mass ($m$) of the u and d quarks
as the third variable, we can study this CP
from the viewpoint of the phase space
$T$--$\mu$--$m$ with a tricritical point (TCP).
The static properties of this phase diagram with the TCP
is well described by the Ginzburg--Landau  (GL) effective potential 
of the quark condensate $\sigma$ expanded up to the $\sigma^6$ term.\cite{LS84}
First in the case of exact chiral symmetry $m=0$
the $T$--$\mu$ plane  must be
devided into two domains of the symmetric and broken phases with a boundary 
{\em line}. Although the order of the singularity
of this line is unknown in general, we {\em assume} the TCP where
the order of the singularity shifts from the 2nd to the 1st order.
Next, when the quark mass takes small but non--zero value,
the 2nd order line disappears and the assumed TCP becomes a usual CP
(see Fig.~\ref{fig:phase}). 
In this consideration, the relation of this CP to the chiral symmetry
is rather obscure.

Characteristic time scale of the system response
becomes infinitely large at a CP, which is known
as critical slowing down. 
The mode whose typical frequency vanishes
at the CP is called ``soft mode.''
In the conventional theory increase of 
this time scale of a soft mode is related to the divergence of 
the susceptibility at the CP.
The most familiar example will be the sigma meson or 
the radial fluctuation of the quark condensate
in the chiral critical transition\cite{HK85}.

In this talk we discuss the soft mode along the
line of the CP within effective approaches to QCD.\cite{HF03,FO04}
We will point out that the soft mode associated with
the CP at finite $m$ must have hydrodynamic character.
This argument is based on 
the conservation of the baryon number and energy densities,
and therefore is very general.
Near the TCP the critical mode is different between the
symmetric and broken phases.
Although we restrict ourselves to the result obtained  
in the Nambu--Jona-Lasinio (NJL) model here, we should stress that
the same result 
can be reached within the time--dependent GL
approach as well.\cite{FO04}

\section{Effective potential}

\begin{figure}[tb]
\centerline{\epsfxsize=4.1in\epsfbox{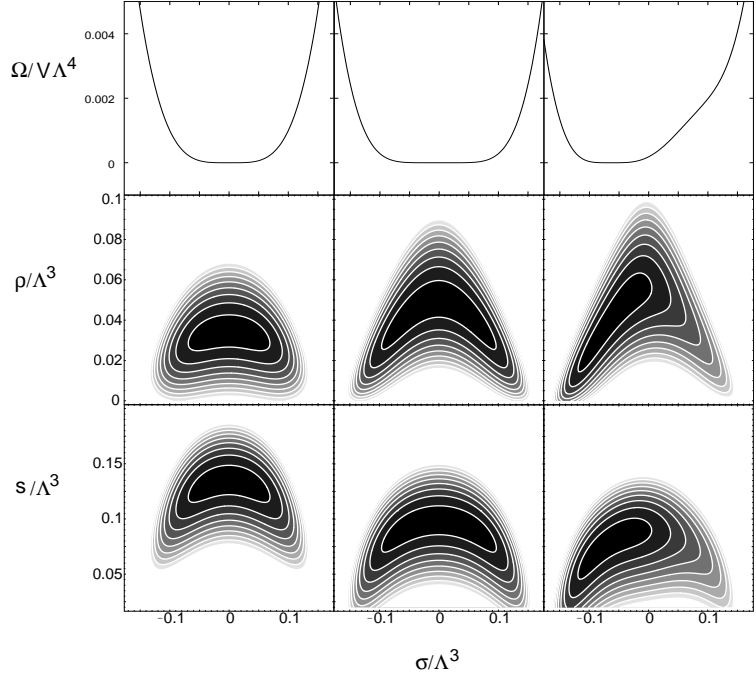}}   
\caption{Effective potentials $\tilde \Omega$ with two ordering densities,
$\sigma$--$\rho$ and $\sigma$--$s$ at critical poinsts,
chiral CP (left), TCP (middle), CP (right).
Those with the single ordering density (uppermost) are
also shown.\label{fig:effpot}}
\end{figure}

We use the NJL model\cite{HK94,SPK92}
$
{\mathcal L}
 = \bar q  (i \!\!\!\not\!\!\partial -m) q +g [
(\bar q q)^2 + ( \bar q i \gamma_5 \tau^a q)^2],
$
in the mean field approximation
($\langle \bar q q \rangle = \sigma$=const, 
$\langle \bar q i \gamma_5 \tau^a q \rangle =\pi$=0).
The thermodynamics is described
by the effective potential,
\begin{eqnarray}
\Omega(T,\mu,m;\sigma)/V&=&
-\nu \int {d^3 k\over (2\pi)^3}
[E - T \ln (1-n_+) -T \ln (1-n_-)]
\nonumber \\
&&+{1\over 4 g}(2g\sigma)^2,
\label{eq:NJLfree}
\end{eqnarray}
where $n_{\pm}=(e^{\beta(E\mp \mu)}+1)^{-1}$, $E=\sqrt{M^2+\bk^2}$,
$M=m-2g \sigma$, and $\nu=2 N_f N_c=2\cdot 2 \cdot 3=12$ with
$N_f$ and $N_c$ the numbers of flavor and color, respectively.
The true thermodynamic state is determined by the 
extremum condition, $\partial \Omega /\partial  \sigma=0$,
and the corresponding grand potential is $\Omega(T,\mu,m)$.
We define the model with the three--momentum cutoff 
$\Lambda$ and with the coupling constant $g \Lambda^2=2.5$
which allows the TCP.
In the following, all the dimensionful 
quantities are expressed in the units of $\Lambda$.

It is useful to difine the GL effective potential $\tilde \Omega$
with two order parameters in studying the true flat direction.\cite{FO04}
It is numerically constructed as shown in Fig.~\ref{fig:effpot}
at the chiral CP with $(T,\mu)$=(0.3419, 0.3),
the TCP (0.20362, 0.49558) and a CP (0.1498, 0.5701) 
with $m=0.01$ in the units of $\Lambda$.\cite{HF03}
The susceptibilities 
$\chi_{ij}=-{1 \over V}\partial ^2 \Omega/\partial i \partial j$
($i,j=T,\mu, m$)
are equal to the inverse of the curvature
matrix of the GL effective potential at the extremum point.
Therefore the divergence of the susceptibilities at 
the critical point is related to the appearance of 
a flat direction in the GL effective potential.

At the chiral CP, the flat direction must be 
the scalar density $\sigma$ due to symmetry
(Fig.~\ref{fig:effpot}).
 The susceptibility of $\sigma$ is divergent while
those of the quark number density $\rho$ and the entropy density $s$ 
remain finite in the mean field approximation.

The $\sigma^2$ and $\sigma^4$ terms in  the potential
(\ref{eq:NJLfree}) vanish at the TCP. 
This fact results in the large fluctuation along the potential valley
of $\tilde \Omega$ with two ordering densities (Fig.~\ref{fig:effpot}). 
The fluctuations of $\rho$ and $s$ become divergent
at the TCP approached from the broken phase.%
\footnote{Note that the fluctuation of $s$ is a linear
 combination of those of the quark number density and the energy density.}
If we define a potential of {\em e.g.,} the
quark density $\rho$ by eliminating $\sigma$ with
$\partial \tilde \Omega(\sigma,\rho)/\partial \sigma=0$,
we see that the potential is flat on the lower density side
of the critical density $\rho_t$ at the TCP.\cite{FO04}

On the other hand, 
at the CP with the explicit breaking $m \ne 0$ 
the proper ordering 
direction becomes a linear combination of
$\sigma$, $\rho$ and~$s$
(Fig.~\ref{fig:effpot}).\cite{KLS01,FO04,SS04}
The susceptibilities of these densities diverge 
with the same exponent
since all of them involve a fraction of 
the critical fluctuation of the proper ordering density.
Here the $\sigma$ direction is no longer special.
One may choose equally well any of these densities as the
ordering density in the static GL potential.

\section{Susceptibility and spectral density}

The susceptibility is obtained in the $q$--limit
of the response function, which
allows us to express the susceptibility as
a sum of the spectral density:
\begin{eqnarray}
\chi_{ij} = \chi_{ij}(0,\bq \to 0)
=\lim_{\bq \to 0}\int \frac{d \omega}{2\pi}
{2 {\rm Im}\chi_{ij}(\omega, \bq) \over \omega}.
\qquad (i,j=m,\mu,T)
\label{eq:sum}
\end{eqnarray}
From this expression we see that divergence of the susceptibility
is caused by spectral enhancement at $\omega =0$ or mode softening, 
provided that the spectral density $2 {\rm Im}\chi_{ij}(\omega, \bq)$ 
itself is integrable. 

At the chiral CP, the divergence of the scalar susceptibility
is generated by softening of the sigma meson mode,
which is the chiral partner of the pion mode.
However, we should recognize that there is no symmetry
reason to expect the massless sigma at the CP with
explicit symmetry breaking due to $m\ne 0$.

There is a strong constraint on the spectrum in the 
fluctuations of the conserved quantity: 
the modes contributing to the susceptibility
of a conserved quantity have to be hydrodynamic, that is,
the typical frequency must vanish as $\bq \to \bzero$.
Physically this is a consequence of the existence of the current $\bf j$
such that $\partial _t \rho+\nabla \cdot {\bf j}=0$ for {\em e.g.,}
the quark number density.
Using the fact that the conserved density operator is commutative with
the total Hamiltonian and the fluctuation--disspation theorem, 
we can express the susceptibility in another form,\cite{HK94,FO04}
\begin{eqnarray}
\chi_{ij}
=\beta \lim_{\bq \to 0}\int \frac{d \omega}{2\pi}
{2 {\rm Im}\chi_{ij}(\omega, \bq) \over 1-e^{-\beta \omega}}.
\label{eq:FDth}
\end{eqnarray}
These two expressions (\ref{eq:sum}) and (\ref{eq:FDth})
coincide with each other if and only if
$\lim_{\bq \to \bzero} 2{\rm Im}\chi_{ij}(\omega,\bq) =
 2\pi  \delta (\omega) \omega \chi_{ij}$.
Hence, when the susceptibility of a conserved quantity diverges,
there must be a hydrodynamic mode which shows critical
slowing.

Finally we remark that the $\omega$--limit of the
response function 
$\chi_{ij}(\omega \to 0, \bzero)$ has no contribution from
the hydrodynamic mode spectrum.

\section{Soft modes in the NJL model}

The response functions in the
random phase approximation are written  as
\begin{eqnarray}
\chi_{ij}(iq_4,{\bf q})
&=&
\Pi_{ij}(iq_4,{\bf q})+\Pi_{i m}(iq_4,{\bf q})
{1 \over 1-2g \Pi_{m m}(iq_4,{\bf q})} 2g \Pi_{m j}(iq_4,{\bf q}).
\nonumber
\\
\label{eq:response}
\end{eqnarray}
Here polarization functions are defined with the 
imaginary--time quark
propagator ${\mathcal S}(\tilde k)=1/(\not \tilde k + M)$ as
\begin{eqnarray}
\Pi_{ij}(iq_4,{\bf q})&=&
-\int {d^3 k \over (2 \pi)^3} T \sum_{n=-\infty}^\infty 
tr_{\rm fcD} {\mathcal S}(\tilde k)\Gamma{\mathcal S}( \tilde k -q)
\Gamma',
\end{eqnarray}
where $q_4=2l\pi T$ $ (l\in {\Bbb Z})$, 
$\tilde k=({\bf k}, k_4+ i \mu)$ with
$k_4=-(2n+1)\pi T$, $\Gamma$ is an appropriate Dirac matrix,
and the traces are taken
over the flavor, color and Dirac indices.
$\Gamma=1$ for the scalar, $i\gamma_4$ for the baryon number,
and ${\mathcal H}_{MF}$ for $\beta$ with
\begin{eqnarray}
{\mathcal H}_{MF}&=&-i\frac{1}{2} {\pmb \gamma}\cdot \!
\stackrel{\leftrightarrow} {\pmb \nabla} + M  + i\mu  \gamma_4.
\end{eqnarray}
We deal with the response function of
 ${\mathcal H}_{MF}$  instead of the entropy
 because the entropy has no microscopic expression.
The real--time response function is obtained 
from the imaginary--time propagator through the usual replacement
$iq_4 \to q_0 +i \epsilon$ in the final expression.

\begin{figure}[tb]
\centerline{\hfill
\epsfxsize=0.45\textwidth\epsffile{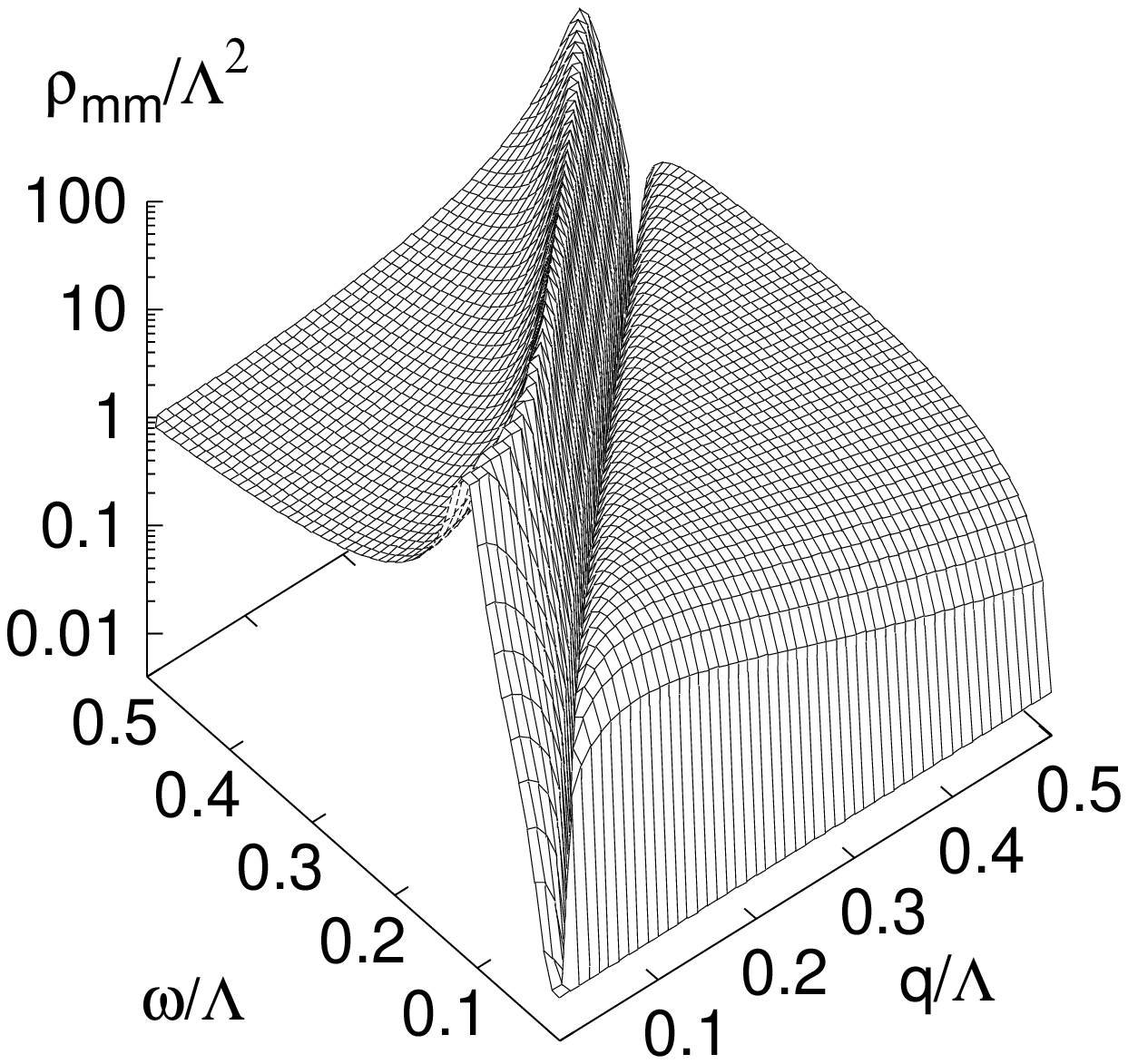}
\hfill
\epsfxsize=0.45\textwidth\epsffile{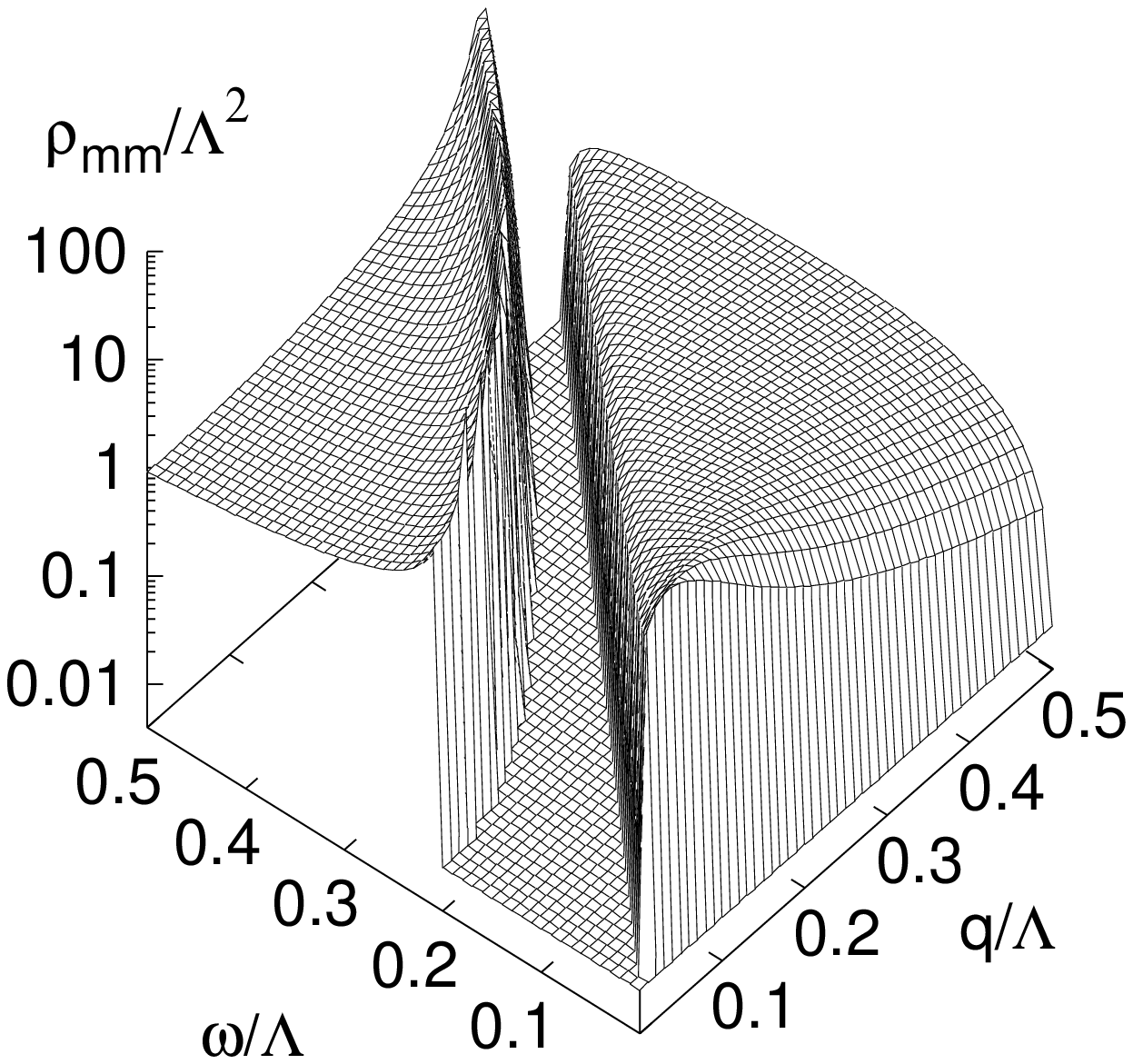}
\hfill}
\centerline{\small \hfill (a) \hfill \hfill (b) \hfill }
\caption{Spectral functions of the scalar channel above (Left, $T/\Lambda=0.35$)
and below (Right, $T/\Lambda=0.339$) the chiral transition point with $\mu/\Lambda =0.3$
in the $\omega$--$q$ plane.}
\label{fig:spfuncchiral}
\end{figure}

\begin{figure}[tb]
\centerline{
\hfill
\epsfxsize=0.35\textwidth
\epsffile{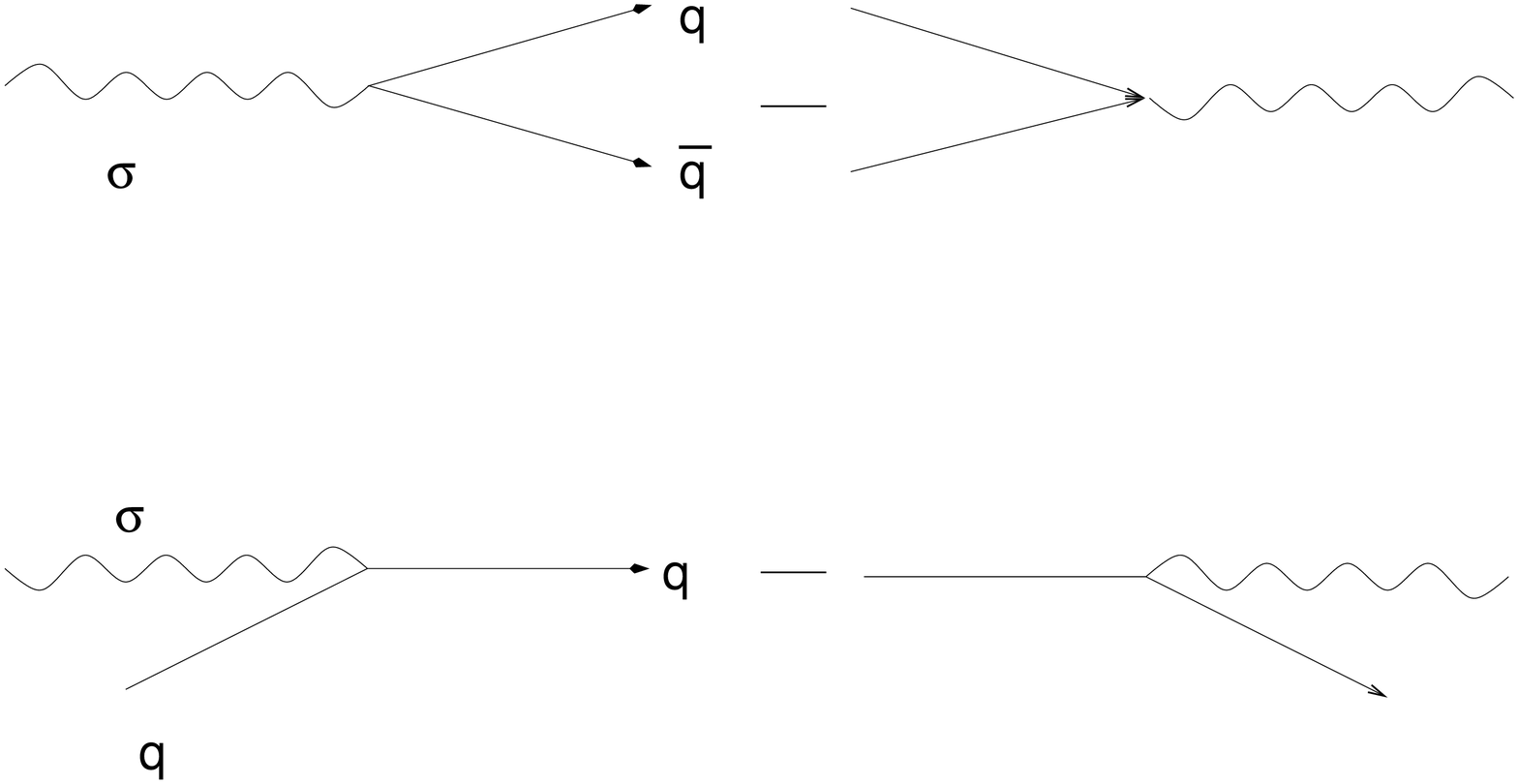}
\hfill
\epsfxsize=0.35\textwidth
\epsffile{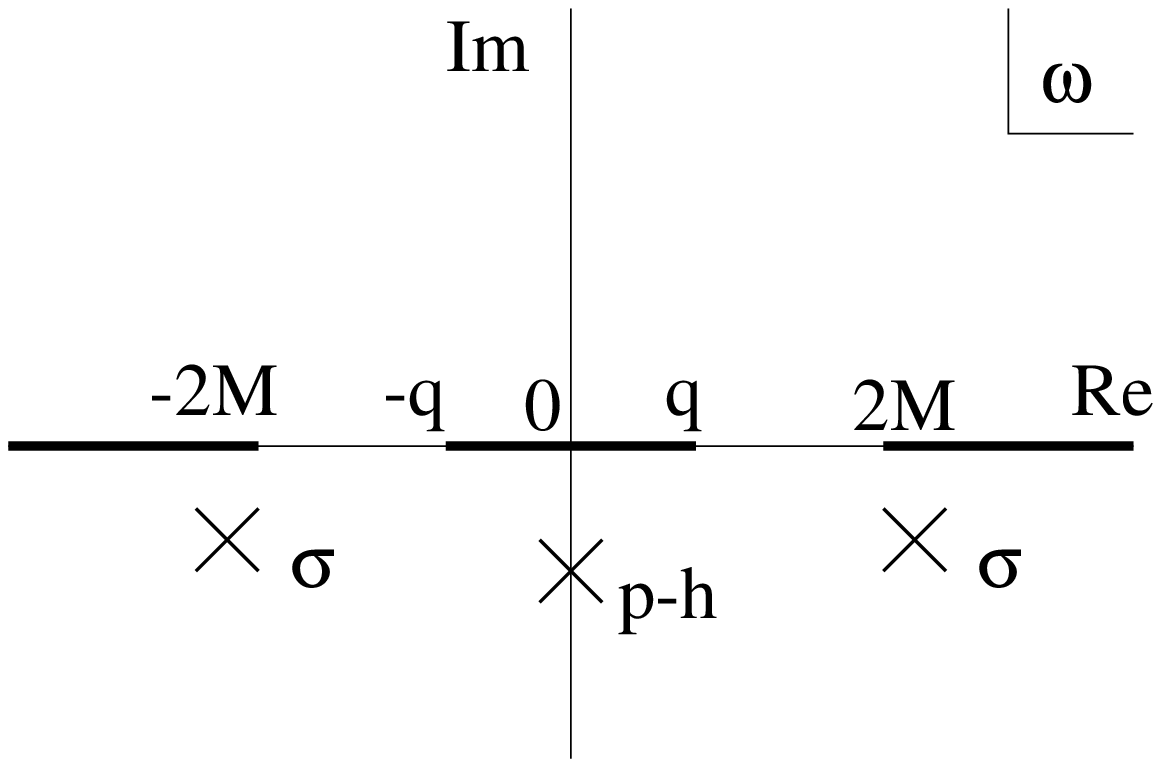}
\hfill
}
\centerline{\small \hfill (a) \hfill\hfill (b) \hfill}
\caption{(a) Physical processes contributing to the spectrum with the detailed 
balance. (b) Analytic structure of the response function.
The $\sigma$ meson  and p--h poles ($\times$)
locate in the unphysical Riemann sheet.
}
\label{fig:cut}
\end{figure}

The collective modes are generated by the
infinite sum of the bubble diagrams in the NJL model. 
The spectral function 
$\rho_{mm}(\omega,\bq)=2 {\rm Im} \chi_{mm}(\omega,\bq)$
 of the scalar response function\cite{HK85} 
just above and below the chiral critical point
with fixed $\mu=\mu_c$ are shown in Fig.~\ref{fig:spfuncchiral}.
The spectrum in the time--like region comes from
the quark pair creation/annihilation while the
spectrum in the space--like region is due to 
the absorption/emission of the scalar fluctuation
by a quark or an anti--quark. The latter 
particle--hole (p--h) process gives rise to
the Landau damping of collective motion in medium. 
These physical processes are schematically
shown in Fig.~\ref{fig:cut} (a).
In the unphyscal Riemann sheet of~$\omega$ (Fig.~\ref{fig:cut} (b)), 
we found two kinds of complex poles corresponding
to the collective excitations of these processes.
We shall call here the pole related with the time--like spectrum
the sigma meson and the pole for the space--like one 
the p--h mode.

Approaching the chiral CP from the symmetric phase,
we see in Fig.~3 (a) that the sigma meson mode becomes soft. 
Meanwhile the p--h mode does not show any particular enhancement.
From the broken phase, on the other hand, the mass gap of the
sigma meson mode is vanishing and the p--h mode spectrum also 
gets stronger at $\bq=\bzero$
(Fig.~3 (b)).

\begin{figure}[tb]
\centerline{\hfill
\epsfxsize=0.47\textwidth\epsffile{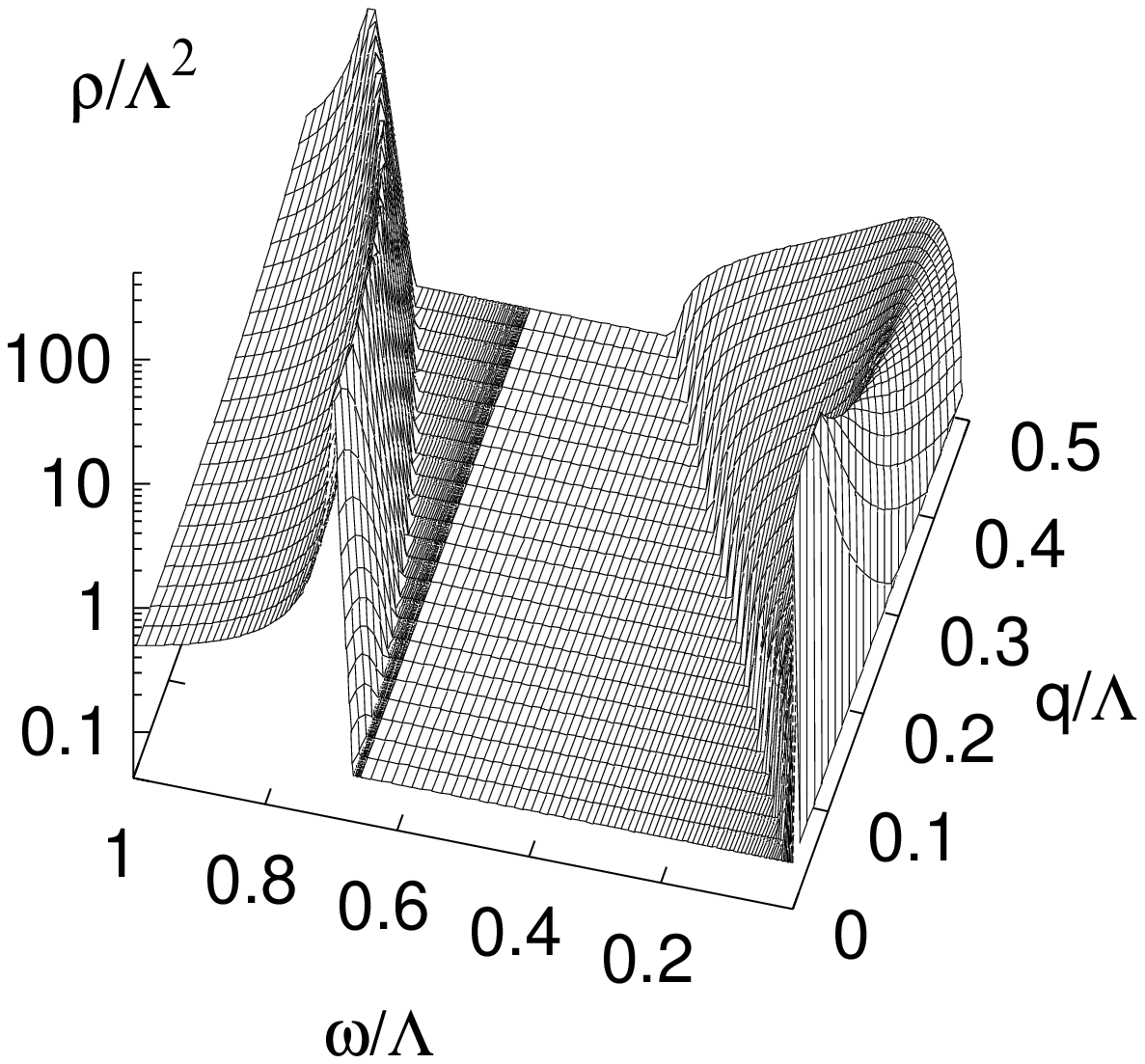}
\hfill
\epsfxsize=0.47\textwidth\epsffile{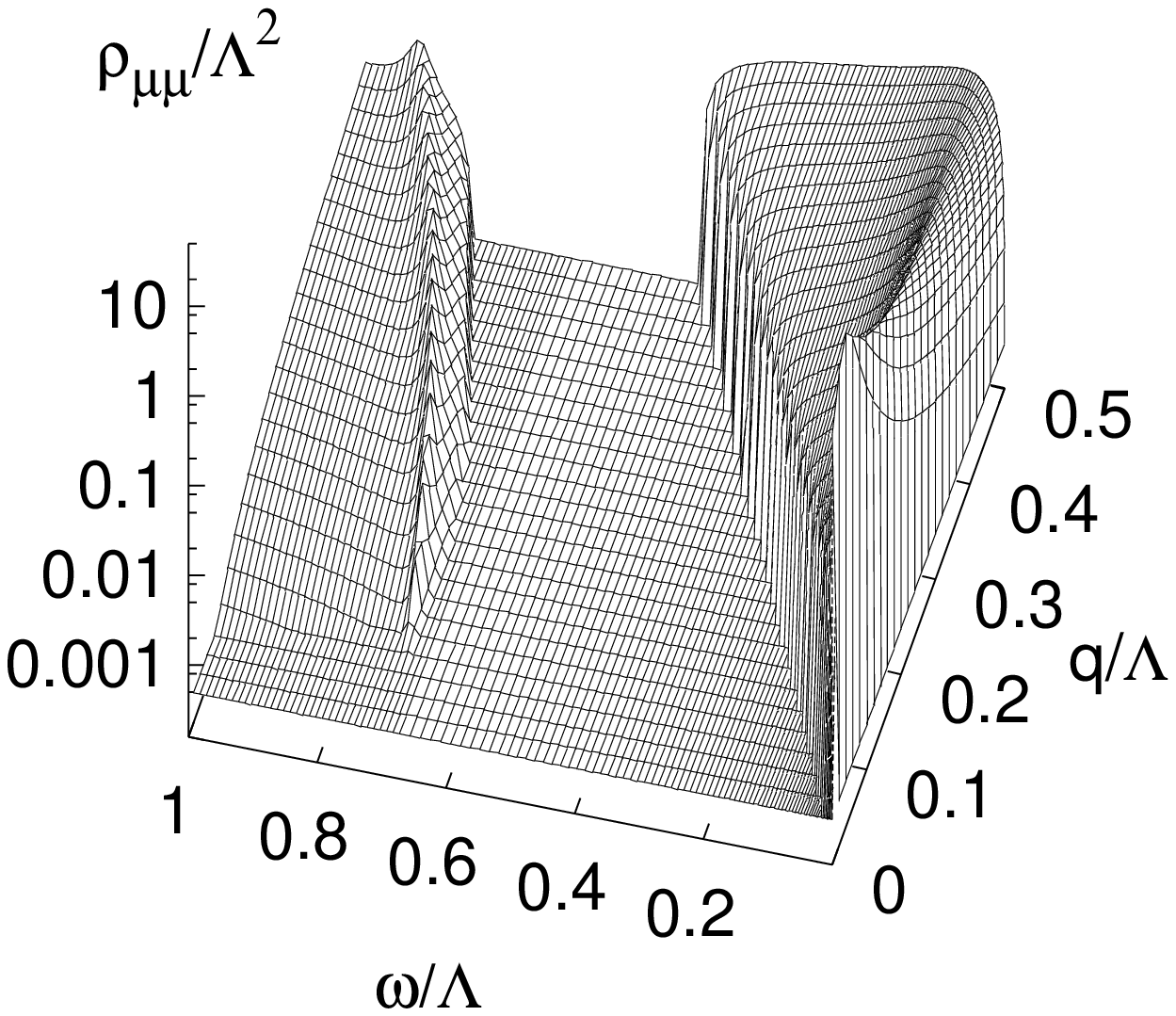}
\hfill}
\centerline{\small \hfill (a) \hfill\hfill (b) \hfill}
\caption{Spectral functions of the scalar and the quark number susceptibilities
at the CP with $m/\Lambda=0.01$.}
\label{fig:spfunc}
\end{figure}

At the critical point with $m/\Lambda =0.01$
both the scalar and quark--number susceptibilities
diverge. 
The scalar spectral function is shown in Fig.~\ref{fig:spfunc} (a).
In this case the sigma meson has
finite mass gap, which is set by the
quark mass as  $\sim 2M \sim m^{1/5}$.
The p--h mode has the hydrodynamic character
($\omega \to 0 $ as $\bq \to \bzero$).
We note that the strength of this mode is strongly enhanced
at this critical point.
In Fig.~\ref{fig:spfunc} (b), we show the specral function
of the quark--number response function. It is clear that the
divergence is caused solely by the p--h mode spectrum with
hydrodynamic character,
which is consistent with the general argument given in Sec.~3.

\section{Change of the spectral contribution along the critial line}

We know that the sigma meson mode is the soft mode associated with the
chiral CP while we have seen that the p--h mode shows
the critical slowing at the CP with finite quark mass $m$.
Let us study the change of the critical eigenmode
along the critical line as shown in a dashed line 
in Fig.~\ref{fig:phase},
with defining the relative weight of the spectral contributions
of the p--h mode to the total spectrum by
\begin{eqnarray}
R
&\equiv&\frac{\chi_{mm}(0,\bzero^+)-\chi_{mm}(0^+,\bzero)}
             {\chi_{mm}(0,\bzero^+)}.
\label{eq:R-NJL}
\end{eqnarray}
Here we used the fact that the difference between the
$q$-- and  $\omega$--limits of
$\chi_{mm}(\omega,\bq)$ 
stems from the spectral contribution of the hydrodynamic mode.
From the argument in Sec.~3, this ratio must be unity
for the susceptibility of a conserved quantity,
which can be confirmed explicitly.

\begin{figure}[tb]
\begin{center}
\epsfxsize=0.45\textwidth 
\epsffile{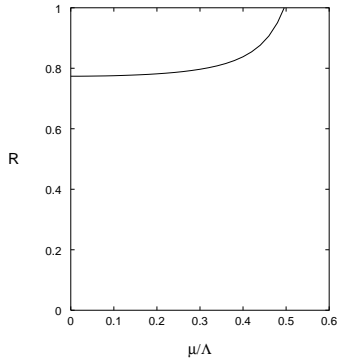}
\vspace{5mm}
\caption{Ratio (\ref{eq:R-NJL}) of the spectral contribution along the chiral
critical line in Fig.~\ref{fig:phase}
approached from the broken phase. $R \to 1$ toward
the TCP. }
\label{fig:ratio}
\end{center}
\end{figure}

We show the numerical result of $R$ in Fig.~\ref{fig:ratio}.
The ratio vanishes if the chiral CP or the TCP is approaced from
the symmetric phase, which means that the critical divergence
is generated by the sigma mode without any contribution
from the p--h mode. From the broken phase, on the other hand,
we see the finite portion of the divergence comes from the
p--h mode spectrum in the chiral critical transition ---
even in the $\mu=0$ case.
Approaching the TCP along the critical line, 
we see that the contribution from the
p--h mode increases and eventually saturates the 
spectral sum~(\ref{eq:sum}).

First one may ask why the p--h mode gives no contibution
in the symmetric phase. This is because that
the chirality and helicity is the same for a massless quark and
that the chirality flip in the scalar coupling requires
the finite momentum transfer $\bq$. In the broken phase, 
a massive quark with definite chirality has both helicity 
components, and therefore the p--h mode spectrum remains contributing
to the susceptibility in the $\bq \to \bzero$ limit.

Next we note that the mixing of this p--h spectrum 
in the susceptibility of $\rho$ and $s$
is possible in the broken phase through the coupling proportional
to the condensate $\sigma$. 
This mixing gives rise to the finite gap of these susceptibilities
across the {\em chiral} critical line. This fact implies that
the p--h mode contribution must be of order of 
$1/\sigma^2 \to \infty$ in the scalar channel,
to cancel the $\sigma^2$ factor from  the coupling.

At the TCP approached from the broken phase with
fixed $\mu$, $\chi_{mm}$ diverges as $1/\sigma^4$ while
$\chi_{\mu\mu}$ and $\chi_{TT}$  blow up as $1/\sigma^2$,
which is easily
confirmed within the Ginzburg--Landau approach.
Only the p--h mode spectrum with hydrodynamic character
can cause this divergence in the NJL model. The sigma
meson mode in the NJL model provides the singularity
of order $1/\sigma^2$ to $\chi_{mm}$ at the TCP approached from the 
broken phase.
From the symmeric phase, where $\sigma \equiv 0$, there is no 
contribution from the p--h mode and the critical 
divergence at the TCP is completely provided by softening
of the sigma meson. The $\chi_{\mu\mu}$ and $\chi_{TT}$ are
finite there.

Along the critical line with finite quark masses $m\ne 0$,
 $\chi_{\mu\mu}$ and $\chi_{TT}$ as well as $\chi_{\mu\mu}$ 
diverge with the same exponent. Generally these divergence 
must come from softening of a hydrodynamic mode in
the system because the baryon number and the energy are
conserved quantities. We can prove this within the
NJL model as well as the time--dependent GL approach.

\section{Summary}
We have seen that, unlike at the chiral CP,
the ordering density at the CP with finite quark mass $m$ is a linear
combination of the scalar density, the baryon number density and the
energy density, and that the susceptibilities of these density
diverge there.
Since the susceptibility of a conserved density solely comes
from the hydrodynamic spectrum, the associated
critical soft mode must be hydrodynamic.
We identified the p--h mode
in the NJL model as this critical mode.
Recently it is explicitly
argued that the dynamic unversality class of this point is the same as
the liquid--gas critical point\cite{SS04,HH77}.

Experimentally, it is worthwhile to study the fluctuation of the
conserved densities.\cite{MAS04}
The correct evolution equation for the correlation length must be
hydrodynamic one, which is slower than
the sigma like motion.
In this sense the growth of the correlation length
in the heavy ion events
passing by the CP acquires a renewed interest.\cite{BR00}

Along the critical line in the phase space of $T$--$\mu$--$m$, we have
studied the changeover of the associated soft mode from the sigma meson mode
at the chiral CP to the p--h mode at the CP with $m\ne 0$.
This shift occurs at the TCP, where the critical soft mode is
different between the symmetric and broken phases.
The dynamic classification of this TCP will be
theoretically interesting.\cite{FO04,SS04,HH77}

There is a speculated phase diagram of QCD in the space of
$T$--$\mu$--$m$--$m_s$,\cite{SA02}
in which we see several critical lines and surfaces.
It seems important to keep the hydrodynamic viewpoint in mind when
we study these criticalities.

\section*{Acknowledgments}

One (H.F.) of the authors is grateful for the warm hospitality
extended to him by KIAS.
This work is supported in part by the Grants-in-Aid for Scientific
Research of Monbu-kagaku-sho (No.~13440067).

%
%
%
%

\end{document}